\documentclass[preprint,pra,nofootinbib]{revtex4}
\usepackage{amscd,fancyhdr}
\usepackage{amsfonts}
\usepackage[mathscr]{euscript}
\usepackage{pstricks,psfragx}
\usepackage{pst-node}
\usepackage{mathrsfs}                                                                %花写字母宏包 格式如：对R字母花写 $\mathscr{H}$
\usepackage{color}                                                                   %支持颜色
\usepackage{float,epsfig}
\usepackage{graphicx}                                                                % Include figure files
\usepackage{dcolumn}                                                                 % Align table columns on decimal point
\usepackage{bm}                                                                      % bold math
\usepackage{amsmath,amssymb,amsthm,mathrsfs}
\usepackage[colorlinks,citecolor=green,linkcolor=blue,anchorcolor=red]{hyperref}    %文献引用的颜色，链接的颜色
\usepackage[stable]{footmisc}
\makeatletter
\def\@hangfrom@section#1#2#3{\@hangfrom{#1#2#3}}
\makeatother
\begin{document}
%%===============================================================================================================================
%%====================================================== The Beginning ==========================================================
%%===============================================================================================================================

%%===============================================================================================================================
\title{A suggested experiment to distinguish between the Bohmian Interpretation and the Standard Quantum Mechanics}
\author{Ke-Xia Jiang$^{a}$\footnote{kexiajiang@126.com; kexiachiang@gmail.com} and  San-Min Ke$^{b,c}$\footnote{ksmingre@163.com}}
\affiliation{$^{a}$Department of Physics, Engineering College of
CAPF, Xi'an 710086, P. R. China
\\
$^{b}$College of Science, Chang'an University, Xi'an 710064, P. R.
China
\\
$^{c}$Key Laboratory for Special Area Highway Engineering of
Ministry of Education, Chang'an University, Xi'an 710064, P. R.
China }
%%===============================================================================================================================
%%===============================================================================================================================
\begin{abstract}
Based on the double-slit experiment of electrons, we suggest a
proposal of thought experiment to distinguish between the Bohmian
Interpretation (BI) and the Standard Quantum Mechanics (SQM). We
mainly focus on the discussion of the meaning of the wave function
(Schr\"{o}dinger-$\psi$). The key technique is require to insert
some slow-electrons or weak electron current into the space between
the double-slit and the detector plane. We find  that the two
theories finally give out two totally different results about the
affections which the externally inserted electrons cause to the
original pattern of the interference fringes. Under the BI, the
externally inserted electrons also can be influenced by the Quantum
Potential (QP) in a totally same way with the electrons which come
from the slits, so the positions they arrived at are preferred to
certain bright zones, and the interference pattern will become more
clearer. While under the SQM, the Schr\"{o}dinger-$\psi$ does not
represent an objectively real field, but only a mathematical
construction of the probability characteristics of the particle
itself, so the externally inserted electrons and the electrons which
come from the slits have no correlations with each other. No any
priority positions at the detector plane the externally inserted
electrons will arrive. And the affections are only the addition of a
uniform bright background. In such a meaning, the dark zones of the
fringes of the interference pattern have been filled.
\end{abstract}
%%===============================================================================================================================
%%===============================================================================================================================
%\pacs{04.70.Dy, 03.65.Xp, 04.62.+v} \documentclass[showpacs]{revtex4}
%%===============================================================================================================================
%%===============================================================================================================================
\date{\today}
%%===============================================================================================================================
\maketitle
%%===============================================================================================================================
%%===============================================================================================================================
\section{Introduction}\label{sec.Introduction}
As we all know, although Quantum Mechanics is one of the pillars of
modern physics, the physical community have not formed a unified
conceptual framework for understanding about the meaning behind the
formulas. Controversies can be traced back to the beginning, while
up to now.

In order to understand conflicts of the SQM with the classical
system, in 1952, D. Bohm\cite{Bohm} suggested another interpretation
using ``hidden variables'', which is occasionally coincided with L.
de Broglie's pilot-wave theory\cite{Broglie}.\footnote{In
literatures, the two theories sometimes calls the de Broglie-Bohm
pilot-wave theory (the dBB). However, since we mainly discuss the
concept \emph{the Quantum Potential} in Bohm's theory, and the dBB
is not used here.} Just as J. S. Bell\cite{Bell1971} pointed out
that there is nothing ``hidden variables'', the main meaning of the
BI is that all variables are forced by the QP. According to the BI,
the wave function represents a new kind of physically real field
that is capable of exerting a force on the particle through its
determination of the QP.

The BI enables one intuitively and imaginatively to understand how
quantum process may actually come about. Maybe as D.
Bohm\cite{Bohm-Hiley1984} has realized that, the approach has not
generally been adopted by the community of physicists, mainly
because it did not lead to any new predictions for the actual
experimental results. However, the indistinguishability with the SQM
is not the inherent character of the BI. As is clearly stated in the
beginning by D. Bohm, effects which caused by modifications of the
suggested interpretation are insignificant in the atomic domain, but
crucial importance in the domain of dimensions of the order of
$10^{-13}$cm or less, where, the Copenhagen Interpretation of the
Quantum Mechanics is totally inadequate.

Suggested experiments that can predict different results for the SQM
and the BI have been the subject of many discussions in literatures
over the years\cite{Brida, Ghose1, Ghose2, Golshani, Ghose3,
Genovese, Akhavan}. Here, we do not analyse and discuss these so
many proposals, and just point out that there are at least three
critical points one should especially pay attention to: (i) At
dimensions of the order of $10^{-13}$cm or more, one may not give
out a suggested proposal. At the state, the BI maybe only provides a
broader conceptual framework that servers as a basis for new or
modified mathematical formulations for the description of physical
system. (ii) Using the particle characteristics of a ``photon"
instead of classical real particles is not proper. According to the
BI, light quanta are described as electromagnetic wave packets which
have many particle-like properties. However, it would not be
consistent to assume the existence of a ``photon'' particle,
associated with each light quantum. (iii) The suggested experiment
should be practically feasible to control and weakly dependent on
energy and environment.

Considering the above attentions, in this paper we tentatively
suggest a proposal of thought experiment to distinguish the two
interpretations of Quantum Mechanics. Our suggested proposal is
based on the double-slit experiment of electrons, which is famous as
a deterministic experiment of its wave-like nature. What we want to
distinguish is: Is the Schr\"{o}dinger-$\psi$ a mathematical
representation of an objectively real field in the BI Or a wave
function only has the probability implication in the SQM which dose
not represent as a real field?
%%===============================================================================================================================
\section{The arrangement setup}\label{sec.setup}
\begin{figure}[t]
\centering \setlength{\unitlength}{1in}
\begin{minipage}[t]{0.8\textwidth}
\centering
\begin{picture}(2.6,2.6)
\put(-0.8,0){\includegraphics[angle=0, totalheight=2.6in]{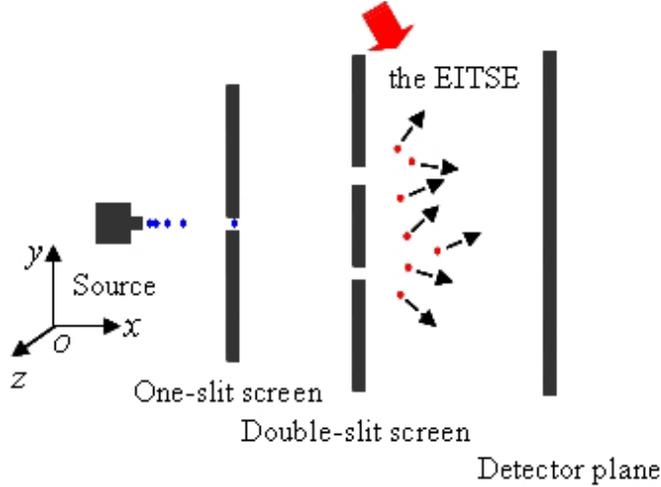}} % 注意使用 Latex 运行时，用的是eps格式的，使用 pdfLatex运行时用的是pdf格式的，故此figures文件夹中有两个格式图片。
\end{picture}
\caption{Schematic Diagram of our proposed thought experiment. The
red dots stand for the EITSE which have been inserted into the space
between the double-slit and the detector plane. The blue dots stand
for electrons emitted form the source which will traverse through
the slits, and finally form a distribution of the interference
pattern if there does not exist the EITSE.\label{fig01}}
\end{minipage}
\end{figure}
The arrangement setup of our proposed thought experiment is not
complicated (see FIG.\ref{fig01}). The key technology is require to
insert some slow electrons or weak electron current with the
velocity in the positive $x$-direction into the space between the
double-slit and the detector plane. These electrons which have low
velocity in the positive $x$-direction come from the outside and do
not from the source. We call them the
externally-inserted-transverse-slow-electrons (abbreviated as the
EITSE). Using the two interpretations, what we want to analysis is
the affections which the EITSE will cause to the interference
pattern formed by electrons from the source (if there does not exist
the EITSE).

%%===============================================================================================================================
\section{The BI predictions about the suggested experiment}\label{sec.BI}
Firstly, let's see what the result is in Bohm's theory. According to
the BI on the double-slit experiment, the form of the QP between the
double-slit and the detector plane pilots the movement behaviors of
electrons. The QP undergoes rapid and violent but harmonic
fluctuations which appear a complex pattern of plateaus and
valleys.\footnote{The form is very reminiscent of the interference
patterns of plane water waves.} This directs a random distribution
of incident particles to form fringes, ever far from the slits. We
can make sure that the form of the QP also has strong influences on
the EITSE instead of only on electrons come from the
source\footnote{We call these electrons the source electrons, and
abbreviate as the SE.}, because in Bohm's theory the
Schr\"{o}dinger-$\psi$ represents an objectively real field and
furtherly electrons should be indistinguishable for the QP. In a
visual imagination, such a form of the QP is similar (but not
identical) with ``trapping wells" for the electrons which moving in
this space wherever they come from. Since the form of the QP can
pilot electrons which coming from the slits to form an interference
pattern, there is no excuse that the QP does not pilot the EITSE in
a same way to give out their contributions to the interference
pattern. So we can sure that the interference pattern will become
more clear (see FIG.\ref{fig0203}(a)). The dark zones of the
interference pattern also have the impossibility for the EITSE under
the effects of the QP. Considering the individual behaviors of
electrons which moving through the space between the double-slit and
the detector plane, it's obviously that this is not in the domain of
the atoms. On the other hand, in Bohm's theory fluctuations of the
QP is under the order of $10^{-13}$cm or less.

The meaning of \emph{slow electrons or weak electron current} is
mainly based on the following expectations. (i) Theoretically, we
hope the EITSE do not change the form of the QP which formed by the
SE. This is reasonable because the QP of the EITSE does not undergo
rapid and violent fluctuations, but is calm. Furthermore, when there
is nothing the EITSE, the form of the QP has successfully piloted
the SE to arrive at the bright zones despite the SE have high
velocity and large energy. (ii) Such a setting guarantees that the
EITSE and the SE will not vastly or directly interact with each
other. (iii) Just considering the non-relativistic case is simple
and enough. (iv) The arrangement should be practically feasible to
control.
%%===============================================================================================================================
\section{The SQM forecast and its differences with the BI predictions}\label{sec.SQM}
Now, let's see what the result is in the SQM forecast. Based on a
simple analysis, we can sure that the SE and the EITSE will not
influence each other in any way. According to the Copenhagen
Interpretation, the Schr\"{o}dinger-$\psi$ describes the electron
itself, while does not represent an objectively real field whose
energy is distributed through space, but only a mathematical
construction with the significance that the intensity in a given
region is a measure for the probability that the particle is
localized there. On the hand, the stochastically inserted EITSE have
no correlations with the SE and further, there are no coherences
between them. So the EITSE and the SE will do not affect (directly
or indirectly) with each other.

\begin{figure}[t]
\begin{minipage}[t]{0.8\textwidth}
\begin{center}
\setlength{\unitlength}{1in}
\begin{picture}(0.4,2)
\put(-2.2,0){\includegraphics[angle=90, totalheight=2in]{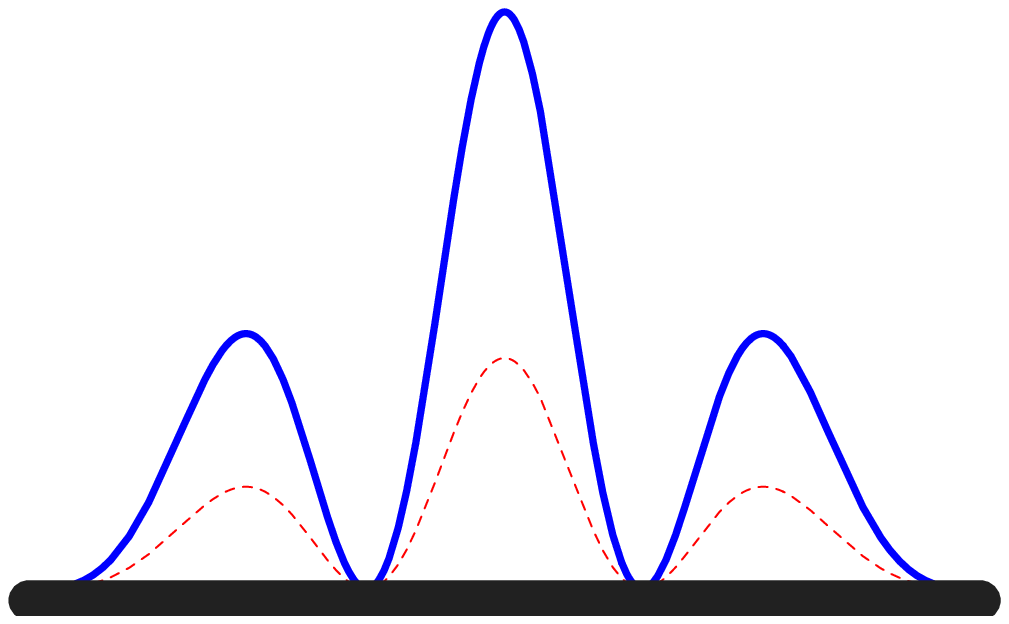}} % 注意使用 Latex 运行时，用的是eps格式的，使用 pdfLatex运行时用的是pdf格式的，故此figures文件夹中有两个格式图片。
\put(0.2,0){\includegraphics[angle=90, totalheight=2in]{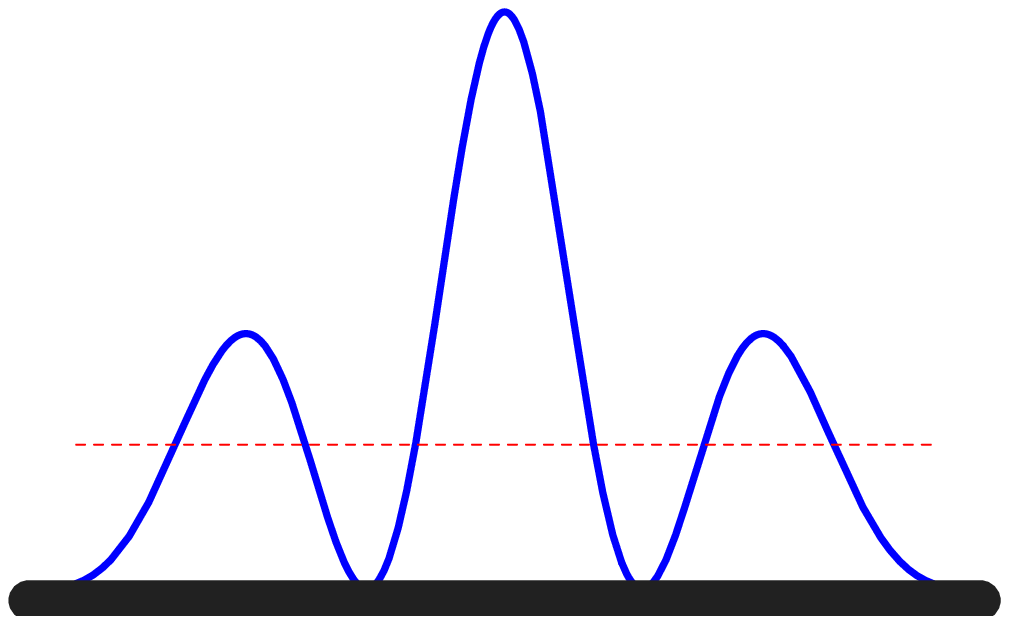}}
\put(-1.4,-0.16){(a)}\put(1,-0.16){(b)}
\end{picture}
\end{center}
\caption{(a)The result under the Bohmian Interpretation. (b)The
result under the Copenhagen Interpretation. The blue solid and red
dashed lines stand for distributions of density of electrons which
from the source and the EITSE, respectively.\label{fig0203}}
\end{minipage}
\end{figure}

If the distribution of the density of the EITSE is uniform in the
positive $x$-direction, and the current is enough, it can be sure
that the EITSE will make the original interference pattern of the SE
becoming fuzzy. Under the original fringes, there will add a uniform
bright background formed by the EITSE. The dark zones of the
original interference pattern will be filled by the bright
background(see FIG.\ref{fig0203}(b)).
%==================================================================================================================
\section{Conclusion and discussion}\label{sec.Conclusion}
In this letter, we suggest a thought experiment in order to
distinguish between the BI and the SQM. The thought experiment has
mainly focused on the discussion of the meaning of the wave function
Schr\"{o}dinger-$\psi$.

With such a arrangement, the two interpretations finally give out
two totally different results about the affections which the EITSE
cause to the fringes of the interference pattern of the SE. In
Bohm's theory, the EITSE also should be influenced by the QP in a
totally same way with the SE. So the positions they arrive at are
also preferred to certain bright zones. While in SQM, the
Schr\"{o}dinger-$\psi$ does not represent an objectively real field,
but only a mathematical construction of the probability
characteristics of the particle itself, so the EITSE and the SE have
no correlations with each other. No any priority positions at the
detector plane the EITSE will arrive. And the affections are only
the addition of a uniform bright background. In such a meaning, the
dark zones of the fringes of the interference pattern have been
filled.

In a same way, the setup of the arrangement can be based on the
other experiments in similar equipment which reveal the wave-like
features of the behavior of microscopic particles, such as
diffraction experiments, and we do not described here detailedly.
The microscopic particles also can be chosen as e.g., neutrons,
protons and atoms.

By use of modern techniques, we have great confidence for the setup
of our proposed arrangement of the thought experiment, because it
feebly depends on energy and environment. The key technology is the
realization of the EITSE. In addition, there is no other
limitations, such as, the initial positions and momentum of the
EITSE. However, the apparatus must also be ingeniously constructed.

%%===============================================================================================================================
\section*{Acknowledgments}
The author(K.-X. Jiang) thanks my dear friend Zi-Wei Ma for kind
help and useful discussions. And S.-M. Ke is supported partially by
the Special Fund for Basic Scientific Research of Central Colleges,
Chang'an University and the Special Foundation for Basic Research
Program of Chang'an University, and also by the Open Fund of Key
Laboratory for Special Area Highway Engineering (Chang'an
University), Ministry of Education (Grant No. CHD2009JC030).
%%===============================================================================================================================


%%===============================================================================================================================

\begin{thebibliography}{99}
\bibitem{Bohm}
D. Bohm, Phys. Rev. 85, 166(1952);\\
D. Bohm, Phys. Rev. 85, 180(1952).

\bibitem{Broglie}
L. de Broglie, \emph{Non-linear Wave Mechanics} (Elsevier:
Amsterdam) (1960).

\bibitem{Bell1971}
J. S. Bell, CERN Preprint TH1424(1971).

\bibitem{Bohm-Hiley1984}
D. Bohm, B. J. Hiley, Found. Phys. 14, 255(1984).

%%===================================================================
\bibitem{Brida}
G. Brida, E. Cagliero, G. Falzetta, M. Genovese1, M. Gramegna and C.
Novero, J. Phys. B: At. Mol. Opt. Phys. 35 (2002) 4751.

\bibitem{Ghose1}
P. Ghose, \emph{Foundations of Quantum Theory and Quantum Optics ed
S M Roy} (Bangalore: Indian Academy of Sciences) (2001) p 211.

\bibitem{Ghose2}
P. Ghose, A. S. Majumdar, S. Guha and J. Sau,  Phys. Lett. A 290
(2001) 205.

\bibitem{Golshani}
M. Golshani and O. Akhavan, 2001 J. Phys. A: Math. Gen. 34 (2001)
5259.

\bibitem{Ghose3}
P. Ghose, Pramana 56 (2001) 211.

\bibitem{Genovese}
M. Genovese et. al., J. Phys. Conf. Ser. 67 (2007) 012047.

\bibitem{Akhavan}
O. Akhavan, \emph{Some Novel Thought Experiments Involving
Foundations of Quantum Mechanics and Quantum Information} (Ph.D
Thesis, Sharif University of Technology) July 2003.


%%===================================================================


\end{thebibliography}
\end{document}